\documentclass[journal]{IEEEtran}
\usepackage{graphicx}
\usepackage{eurosym}
\usepackage{microtype}
\usepackage{amssymb}
\usepackage{yfonts}
\usepackage{url}
\usepackage{multirow}
\usepackage{xcolor}
\usepackage{blindtext}
\usepackage{eucal}
\usepackage{enumitem}
\usepackage{amsmath}
\usepackage{algorithmic}
\usepackage{nomencl}
\usepackage{etoolbox}
\usepackage{eurosym}
\usepackage{amsfonts}
\usepackage{array}
\usepackage{comment}

\usepackage[caption=false]{subfig}

\makenomenclature

\renewcommand\nomgroup[1]{%
\color{black}
  \item[\bfseries
  \ifstrequal{#1}{P}{Variables}{%
  \ifstrequal{#1}{C}{Constants}{%
  \ifstrequal{#1}{S}{Sets and Indices}{}}}%
]\vspace{7pt}}

\makeatletter
\newcommand{\thickhline}{%
    \noalign {\ifnum 0=`}\fi \hrule height 0.75pt
    \futurelet \reserved@a \@xhline
}
\newcolumntype{"}{@{\hskip\tabcolsep\vrule width 0.75pt\hskip\tabcolsep}}
\makeatother

\begin{document}

\title{
Power to Air-transportation via Hydrogen}
\author{Alireza~Soroudi, Soheil~Jafari
\thanks{
Alireza Soroudi (alireza.soroudi@ucd.ie) is with the School of Electrical and Electronic Engineering, University College Dublin, Dublin 04, Ireland. The work of Alireza Soroudi has emanated from research
supported (in part) by University College Dublin under the UCD Grant Number OBRSS-62184.
Soheil Jafari (S.Jafari@cranfield.ac.uk) 
is with the Centre for Propulsion Engineering, School of Aerospace, Transport and Manufacturing (SATM), Cranfield University, Cranfield, UK.
}
}

\markboth{}%
{Shell \MakeLowercase{\textit{et al.}}: Bare Demo of IEEEtran.cls for Journals}
\maketitle

\begin{abstract}
This paper proposes a framework to analyze the concept of power to hydrogen (P2H) for fueling the next generation of aircraft. The impact of introducing new P2H loads is investigated from different aspects namely, cost, \textcolor{black}{carbon emission}, and wind curtailment. The newly introduced electric load is calculated based on the idea of replacing the busiest international flight route in the Europe, Dublin-London Heathrow, by hydrogen fuel-powered aircraft as a high potential candidate for the next generation of air travel systems to cope with the ambitious targets set in Europe Flight Path 2050 by the Advisory Council for Aeronautics Research in Europe (ACARE). The simulation is performed on a representative Irish transmission network to demonstrate the effectiveness of the proposed solution. 
\end{abstract}
\IEEEpeerreviewmaketitle

\nomenclature[P, 01]{$C_f$}{Fuel cost}
\nomenclature[P, 01]{$C_e$}{Emission cost}
\nomenclature[P, 01]{$C_{lsh}$}{Load shedding cost}
\nomenclature[P, 01]{$C_{wc}$}{Wind curtailment cost}
\nomenclature[P, 01]{$C_{P2H}$}{Investment cost for P2H plant}
\nomenclature[C, 01]{$\tau_t$}{Duration of period t (hours)}
\nomenclature[C, 01]{$\lambda_e$}{Emission cost coefficient}
\nomenclature[C, 01]{$\lambda_D$}{load shedding cost coefficient}
\nomenclature[C, 01]{$\lambda_W$}{Wind curtailment cost coefficient}
\nomenclature[P, 01]{$c_{g,t}$}{Fuel cost of unit g at time t}
\nomenclature[P, 01]{$P^{SH}_{b,t}$}{Active load shedding at bus b and time t}
\nomenclature[P, 01]{$\xi$}{Capacity of P2H plant (MW)}
\nomenclature[C, 01]{$\Xi$}{Price of investment for P2H plant (\euro/MW)}
\nomenclature[P, 01]{$P^C_{b,t}$}{Wind curtailment at bus b and time t}
\nomenclature[P, 01]{$P^G_{g,t}$}{Power of thermal generator g at time t}
\nomenclature[P, 01]{$P^W_{b,t}$}{Power of wind turbine in bus b at time t}
\nomenclature[C, 01]{$P^L_{b,t}$}{Electric demand in bus b at time t}
\nomenclature[P, 01]{$Pch_{t}$}{Extracted power from the grid to produce $H_2$ at t}
\nomenclature[P, 02]{$Pdch_{t}$}{Discharged power from $H_2$ storage at t}
\nomenclature[P, 01]{$SOC_{t}$}{State of charge in $H_2$ storage at t}
\nomenclature[C, 01]{$\zeta_{b}$}{0/1 parameter indicating if P2H plant exists in bus b}
\nomenclature[P, 01]{$P^\ell_{bi,t}$}{Power flow in line $\ell$ at time t}
\nomenclature[C, 01]{$a_g,b_g,c_g$}{Thermal power cost coefficients}
\nomenclature[C, 01]{$B_{bi}$}{Susceptance of line between bus b and i}
\nomenclature[C, 01]{$RU_{g}$}{Ramp up of thermal generator g}
\nomenclature[C, 01]{$RD_{g}$}{Ramp down of thermal generator g}
\nomenclature[P, 01]{$\delta_{i,t}$}{Angle of bus i at time t}
\nomenclature[C, 01]{$D_{H_2}$}{Daily hydrogen demand}
\nomenclature[C, 01]{$D_f$}{Daily aircraft fuel burn}
\nomenclature[C, 01]{$N_s$}{Average seats per aircraft}
\nomenclature[C, 01]{$FHV$}{Fuel Heat Value}

\nomenclature[C, 01]{$P_{ef}$}{\textcolor{black}{Equivalent jet fuel cost (\euro/MWh)}
\nomenclature[C, 01]{$P_{cos}$}{\textcolor{black}{Carbon offsetting cost}}
\nomenclature[C, 01]{$P_{f}$}{\textcolor{black}{Jet fuel cost}}}

\nomenclature[C, 01]{$\Lambda^W_{b}$}{Capacity of wind turbine in bus b}
\nomenclature[C, 01]{$w_t$}{Wind power availability at time t}
\nomenclature[C, 01]{$N_f$}{Number of flights per day}
\nomenclature[C, 01]{$N_s$}{Average number of seats per aircraft}
\nomenclature[C, 01]{$\gamma$}{Aircraft fuel burn per journey}
\nomenclature[C, 01]{$d_t$}{Demand at time t}
\nomenclature[S, 01]{$\Omega_G$}{Set of thermal generating units}
\nomenclature[S, 01]{$\Omega_B$}{Set of network buses}
\nomenclature[S, 01]{$\Omega_L$}{Set of transmission lines}
\nomenclature[S, 01]{$\Omega_D$}{Set of time steps in day D}
\nomenclature[S, 01]{$\Omega_k$}{Set of segments for linearizing the cost function}

\vspace{-1mm}
\printnomenclature

\vspace{-3pt}
\section{Introduction}\label{SecI} 
The increased share of renewable energy resources specially non-synchronous technologies such as wind and solar power can create new technical challenges for the power system operators. These challenges include (but not limited to) uncertainty in generation output of RES technologies \cite{mohseni2017optimal}, stability issues due to low inertia \cite{kheradmandi2017using} and voltage control requirements. The inertia problem is prominent for those countries which are weakly connected to their neighbors with AC interconnection links. As an example, a huge amount of wind is annually curtailed in Ireland due to several technical reasons such as transmission network constraints or stability issues. In 2018, the total wind dispatch down in republic of Ireland and Northern Ireland was 707 GWh \cite{eirgrid2018annual}.
The three main reasons are listed as follows:
\begin{itemize}
    \item The transmission constraints are related to the thermal limits of transmission lines as well as the N-1 security requirements. Due to the geographical availability of the wind, most wind turbines are connected to the west and south-west of the country and this increases the chance of wind dispatch-down in Ireland.
    \item By increasing the penetration level of non-synchronous generations (such as wind turbines), the amount of inject-able wind power to the grid is limited to avoid frequency control problems. The total on-line inertia in the system should be able to provide rapid frequency response in case of a disturbance. \textcolor{black}{To keep the power system safe and
    secure it is vital to measure and limit the System Non-Synchronous Penetration (SNSP) \cite{kuwahata2020renewables}.}
    \item Even if the amount of available on-line inertia is sufficient, the ramp-rates of thermal generating units might be not enough to cope with the variations of wind generation and cause wind dispatch-down. The min power generation limit in conventional technologies is another reason for curtailing the available wind power. 
\end{itemize}
\textcolor{black}{
\begin{table*}[ht]
	\renewcommand{\arraystretch}{1}
	\caption{\textcolor{black}{Literature review of some existing works}}
	\label{tab:LR}
	\centering
	\scalebox{1}{
	\begin{tabular}{c|ccc|cc|cccc}
		\hline
Reference	&	Ship	&	Vehicle 	&	Air transportation 	&	Power flow constraints 	&	SNSP	&	Device based 	&	System 	&	Distribution/Transmission 	\\ \hline
\cite{9018500}	&		&	 \checkmark	&		&	No	&	No	&		&	Yes	&	NA	\\
\cite{9069829}	&		&	 \checkmark	&		&	AC	&	No	&		&	Yes	&	Distribution small scale 	\\
\cite{8932711}	&	        	&	 \checkmark	&		&	No	&	No	&		&	Yes	&	Distribution small scale 	\\
\cite{7473861}	&       		&	 \checkmark	&		&	No	&	No	&	Yes	&		&	NA	\\
\cite{9018494}	&	            	&	General	&	    	&	Yes	&	No	&		&	Yes	&	Transmission\\
\cite{8370335}	&		            &	General	&		    &	No	&	No	&		&	Yes	&	Distribution small scale 	\\
\cite{9018493}	&		            &	General	&		    &	No	&	No	&	Yes	&		&	NA	\\
\cite{4397074}	&		            &	General	&		    &	No	&	No	&	Yes	&		&	NA	\\
\cite{8976275}	&	 \checkmark	    &		    &	    	&	No	&	No	&	No	&	No	&	NA	\\
\cite{mcdonagh2020hydrogen}	&		&	General	&		    &	No	&	No	&	Yes	&		&	NA	\\
Proposed model	&		&		&	\checkmark	&	DC-OPF	    &	Yes	&		&	Yes	&	Transmission-Large scale	\\
\hline
	\end{tabular}}
\end{table*}
}

Dispatching down the clean energy (and using fossil fuel based technologies) is not in line with the European goals for reducing the  greenhouse gas emissions by at least 40\% from 1990 levels \cite{NECPs}. Some approaches have been proposed in the literature to reduce the wind curtailment. These methods can be categorised into two main groups:
\begin{itemize}
    \item Non-wire solutions: in these models, the system operator/regulator tries to use the existing assets and utilise them in a more efficient way \cite{soroudi2017resiliency}. For example, dynamic line rating \cite{morozovska2020dynamic}, using distributed series power flow controllers \cite{soroudi2017resiliency} and novel control techniques \cite{djordjevic2019mathematical}. 
    \item Asset building models: in these models, some transmission lines will be built or upgraded \cite{maghouli2016transmission}. Investing in energy storage technologies \cite{yacar2018storage} can be included in this group. 
\end{itemize}
Building new assets has been the traditional way of answering technical challenges in energy sector. However, the difficulties in obtaining the public acceptance is a big challenge for the decision makers in energy sector. In this work, the focus is on improving the principals that the power system is operated based on them without the need for upgrading the transmission network. \textcolor{black}{The $H_2$ production has been proposed in the literature to overcome mitigate
sub-synchronous oscillation in wind power
systems \cite{8964064} and seasonal storage \cite{GABRIELLI2020109629}. Some existing works in the literature are compared in Table \ref{tab:LR}.
The idea of using zero-emission Hydrogen ($H_2$) as an alternative for fossil fuels in transportation sector has received a great deal of attention recently. 
One of the main drivers for this transition is EU’s energy and climate targets for 2030. 
}
\subsection{Hydrogen based airplanes}
\begin{figure}[th]
  \centering
	\includegraphics[width=0.45\columnwidth,bb=140 100 470 750,angle=90]{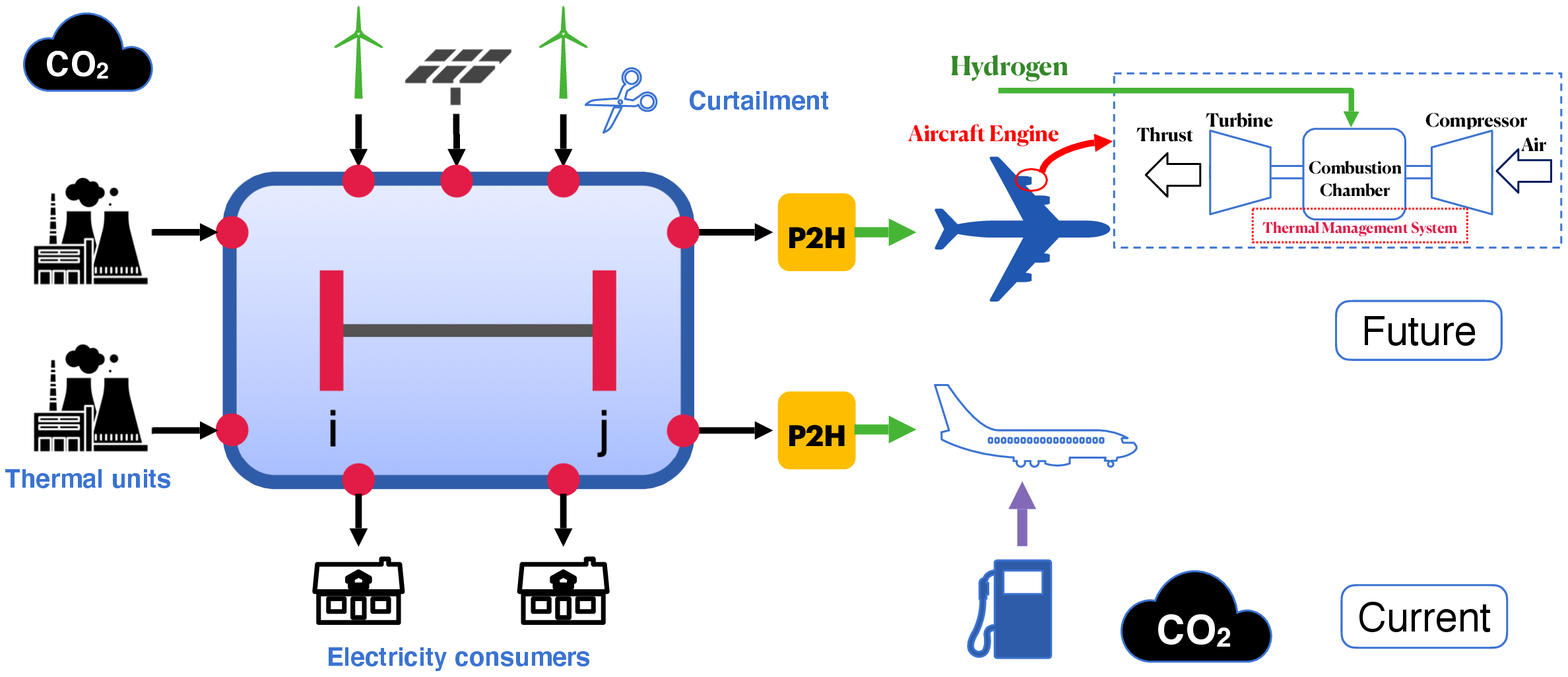}
  \caption{Framework for future hydrogen based air-transportation}
\label{fig:config}
\end{figure}

The aviation sector is one the fastest-growing polluters. forecasts claim that aviation emissions could double in the next three decades even with more fuel-efficient aircraft \cite{aviationmanchester}. So, it is the time to think differently about revolutionary ideas to deal with this challenge.
The idea of liquid hydrogen fuel-powered aircraft has been widely considered in recent years by well-known airliners and manufacturers \cite{concept}.\\ \indent Technical feasibility, safety, environmental compatibility, and economic viability of liquid hydrogen as an alternative fuel for the next generation of civil aircraft have been confirmed by European Union (EU) researchers from 34 institutions in collaboration with industrial partners like Airbus \cite{Airbus34}. The energy density of hydrogen per unit mass is 2.8 times higher than traditional jet fuel. However, the main obstacle is to produce it in quantity from low-carbon energy sources such as wind or nuclear. If this bottle neck is addressed, the aircraft greenhouse gases will be reduced dramatically. This would be a game changer for aircraft manufacturers to meet the severe limitations set in Advisory Council for Aviation Research and Innovation in Europe (ACARE) Flight Path 2050 (e.g. a 75 percent reduction in $CO_2$ emissions and a 90 percent reduction of $NO_x$ emissions by 2050) \cite{ACARE}.\\ \indent The other advantage of using liquid hydrogen in aviation is its thermal management characteristics as one of the best coolants used in engineering applications. The design of new and next generation of aircraft engines are increasingly complex with higher demands on engines for thrust and power generation resulting in hotter fluids, higher components temperature and higher heat generation, which means critical thermal management issues. So, it is time to think differently about how thermal loads in modern gas turbine engines can be managed. The liquid hydrogen is a very high potential heat sink that could be used in the architecture of thermal management systems for aircraft propulsion \cite{soheilTMS}. 
\textcolor{black}{
This paper proposes the idea of converting green electricity to hydrogen and supply the fuel need for the next generation of airplanes as shown in Fig. \ref{fig:config}. The research gaps are as follows:\\
\begin{itemize}
    \item The SNSP impact on $H_2$ production is not properly investigated
    \item The energy requirements of future hydrogen based airplanes and their impacts on power systems should to be analysed.  
\end{itemize}
} 

\textcolor{black}{The following questions will be answered:
\begin{itemize}
    \item How much energy do the next generation of airplanes need? 
    \item How capable is the power system to supply those energy needs via green renewable resources? 
    \item How power system-Air transportation nexus is achievable using the existing assets?
\end{itemize}}
\subsection{Contributions}
An energy procurement model is proposed and formulated with the following properties:
\begin{itemize}
    \item Proposing a linear model for procuring $H_2$ required for Air-transportation
    \item Considering the SNSP constraint for a system with high penetration level of non-synchronous generation
    \item Modeling the environmental impacts of aviation and electricity nexus.
    \item \textcolor{black}{Modeling flight envelope's energy requirement.} 
\end{itemize}
\subsection{Paper structure}
The remainder of this paper is organized as follows: Section \ref{sec:pf} presents the proposed model, as well as the main objective function and related constraints. In Section \ref{sec:simulation} 
the simulations are carried out on the Irish representative system and the numerical investigations are presented. Finally, the paper concludes in Section \ref{sec:Conclusion}.

\section{Problem formulation}
\label{sec:pf}
This section of the paper provides the proposed formulation with the underlying assumptions. 
\subsection{Objective function}
Objective function of the proposed algorithm includes operational costs, load shedding costs, wind curtailment cost and the annualized investment costs of P2H plants. The costs defined as the total  costs should be minimized: 
\begin{alignat}{2}
\label{eq:ofnormal}
OF=C_{f}+C_{e}+C_{lsh}+C_{wc}+C_{P2H}
\end{alignat}
The different costs in (\ref{eq:ofnormal}) are defined as follows:
\begin{alignat}{2}
\label{eq:costs}
&C_{f}=\sum_{g,t} \tau_t C_{g,t} \\
&C_{e}=\sum_{g,t} \tau_t \lambda_e P^G_{g,t} \\
&C_{lsh}=\sum_{b,t} \tau_t\lambda_D P^{SH}_{b,t}\\
&C_{wc}=\sum_{b,t} \tau_t\lambda_W P^C_{b,t}\\
& C_{P2H}=\xi \times \Xi
\end{alignat}
where $C_{g,t} $ is the fuel cost function ( \euro/h), which is modeled by a quadratic function as follows.
\begin{alignat}{2}
\label{eq:Fp}
& C_{g,t}=a_g(P^G_{g,t})^2+b_gP^G_{g,t}+c_g
\end{alignat}
The non-linear fuel cost function in (\ref{eq:Fp}) will be replaced by set of equations as described in (\ref{eq:DEDLP}) \cite{Soroudi2017}:
\textcolor[rgb]{0,0,0}{
\begin{subequations}
\label{eq:DEDLP}
\begin{alignat}{2}
\label{eq:DEDLPa}
&C_f=\sum_{g,t} C_{g,t}\\
\label{eq:DEDLPf}
& C_{g,t}=a_g(P^{min}_g)^2+b_gP^{min}_g+c_g+\sum_k s^k_g P^k_{g,t} \\
\label{eq:DEDLPg}
& s^k_{g}= \frac{C^k_{g,fin}-C^k_{g,ini}}{\Delta P^k_g} \\
& C^k_{g,ini}=a_g(P^k_{g,ini})^2+b_gP^k_{g,ini}+c_g \\
& C^k_{g,fin}=a_g(P^k_{g,fin})^2+b_gP^k_{g,fin}+c_g \\
\label{eq:DEDLPb}
& 0 \leq p^k_{g,t} \leq \Delta P^k_g, \forall k \in \Omega_k\\
\label{eq:DEDLPc}
& \Delta P^k_g = \frac{P^{max}_g-P^{min}_g}{|\Omega_k|}\\
\label{eq:DEDLPd}
& P^k_{g,ini}=(k-1) \Delta P^k_g + P^{min}_g \\
& P^k_{g,fin}= \Delta P^k_g + P^k_{g,ini} \\
\label{eq:DEDLPe}
& P^G_{g,t}= P^{min}_g+\sum_k P^k_{g,t} 
\end{alignat}
\end{subequations}
}
\textcolor{black}{The proposed linear formulation of the operating cost makes the problem scalable and suitable for realistic large power systems.}

\subsection{Constraints}
The DC power flow equations and constraints of the system are as follows ($\forall t \in \Omega_T, \forall b,i \in \Omega_B$) :

\textcolor{black}{The nodal electric power balance for each time period is described as:}
\begin{alignat}{2}
\label{eq:dcpf1}
&\sum_{g=1}^{\Omega_{G_b}}P^G_{{g,t}}+P^W_{b,t}-P^L_{b,t}+P^{SH}_{b,t}-Pch_t\times \zeta_b=\sum_{i=1}^{\Omega_B} P^\ell_{bi,t}
\end{alignat}
\textcolor{black}{The power flow between each pair of connected buses (i.e. $b-i$) is calculated as:}
\begin{alignat}{2}
\label{eq:flow}
&P^\ell_{bi,t}=B_{bi}(\delta_{b,t}-\delta_{i,t})
\end{alignat}
\textcolor{black}{The satisfaction of ramp rate limits of the thermal units are ensured via the following constraints:}
\begin{alignat}{2}
&P^G_{g,t}-P^G_{g,t-1} \leq RU_g \\
&P^G_{g,t-1}-P^G_{g,t} \leq RD_g 
\end{alignat}
\textcolor{black}{The hourly state of charge for hydrogen storage is dependant on the hourly charge/discharge as well as the efficiency factors.}
\begin{alignat}{2}
& SOC_t=SOC_{t-1}+(Pch_t-Pdch_t)\tau_t
\end{alignat}
\textcolor{black}{The amount of hydrogen discharge depends on the capacity of the P2H plant ($\xi$).}
\begin{alignat}{2}
& Pdch_t \leq \xi 
\end{alignat}
\textcolor{black}{The total daily discharged hydrogen should be greater than the daily hydrogen demand for transportation ($D_{H_2}$).} 
\begin{alignat}{2}
& \sum_{t \in \Omega_D} Pdch_t \geq D_{H_2} 
\end{alignat}
where the following operating limits should be considered $\forall t \in \Omega_T$.
\textcolor{black}{The power output of thermal generating units should be kept within min and max values.}
\begin{alignat}{2}
\label{eq:dcpf2}
& P_{g}^{min} \leq P^G_{g,t} \leq  P_{g}^{max} \ \ \ \  \ && \forall g \in \Omega_G
\end{alignat}

\textcolor{black}{The voltage angles of each bus should be kept within min and max values.}
\begin{alignat}{2}
\label{eq:dcpf4}
& \delta_{b}^{min}\leq \delta_{b,t} \leq \delta_{b}^{max}  && \forall b \in \Omega_B
\end{alignat}
\textcolor{black}{The power flow on each transmission line should be less than the thermal limit of that line.}
\begin{alignat}{2}
\label{eq:dcpf5}
& -\bar{P}_{bi}^{\ell} \leq P^\ell_{bi,t}\leq \bar{P}_{bi}^{\ell}  && \forall \ell \in \Omega_L
\end{alignat}
\textcolor{black}{The power output of wind turbines ($P^W_{b,t}$) are dependant on the wind availability of the region. The curtailed wind power is also limited to the wind availability and installed wind capacity.}
\begin{alignat}{2}
&0 \leq P^W_{b,t} \leq w_t \Lambda^W_{b} && \forall b \in \Omega_B\\
& 0 \leq P^C_{b,t} \leq  w_t \Lambda^W_{b}-P^W_{b,t} && \forall b \in \Omega_B 
\end{alignat}
\textcolor{black}{The variations of hourly electric demand (without the hydrogen demand) are assumed to be known as the input data: }
\begin{alignat}{2}
& P^L_{b,t} = d_t\bar{P}^L_{b} && \forall b \in \Omega_B 
\end{alignat}
The SNSP limit is enforced as follows:
\begin{alignat}{2}
\label{eq:SNSP}
& \frac{\sum_bP^W_{b,t}+P^{import}_{t}}{\sum_bP^L_{b,t}+Pch_t+P^{export}_{t} } \leq SNSP
\end{alignat}
\textcolor{black}{Considering the SNSP limit ensures the stability of the power system specially in the case of high wind power penetration.}
\subsection{Decision variables and input parameters}
The decision variables are listed here:
\begin{alignat}{2}
&DV=\Bigg\{\begin{matrix}
P^G_{g,t}, P^{SH}_{b,t},P^W_{b,t}, P^C_{b,t}\\
\xi, Pch_t,Pdch_t,SOC_t\\
P^\ell_{bi,t},\delta_{b,t}
\end{matrix} \Bigg\}
\end{alignat}
The input data for this problem are listed here:
\begin{alignat}{2}
&Data=\Bigg\{\begin{matrix}
[a,b,c]_{g}, \lambda_e, \lambda_D, \lambda_W, \tau_t \\
B_{bi},RU_g,RD_g,P^{min}_g,P^{max}_g \\
d_t,w_t,\Lambda_b^W, \bar{P}^L_{b},D_{H_2} 
\end{matrix} \Bigg\}
\end{alignat}

\section{Simulation results}
\label{sec:simulation}
The proposed formulation is coded in GAMS \cite{Soroudi2017} and the simulation results are discussed in this section. 
\subsection{Aircraft data}
\begin{figure*}[t!]
  \centering
	\includegraphics[width=0.123\columnwidth,bb=298 260 347 550]{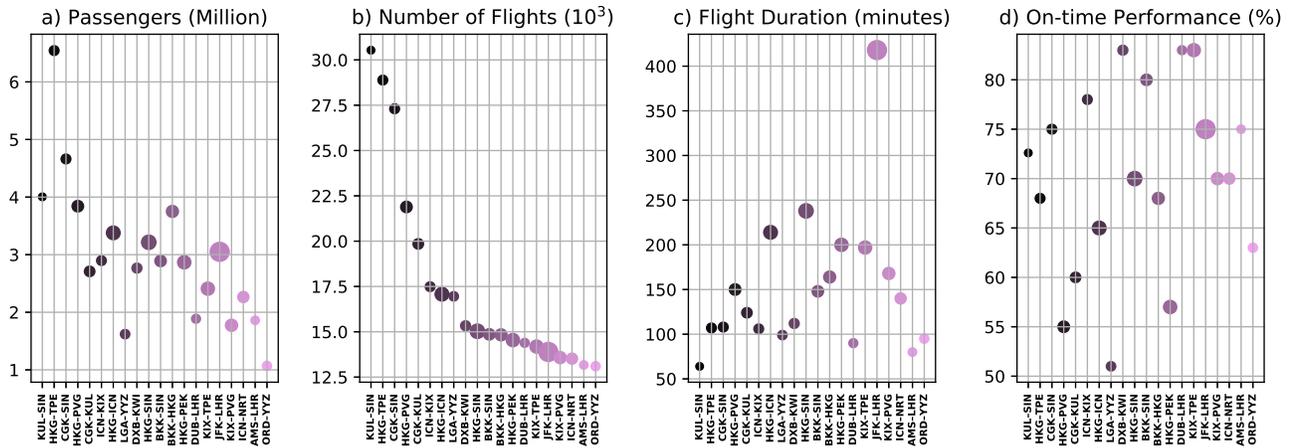}
  \caption{{Statistical data for top 20 busiest international air routes in the world, a) Number of passengers b) Number of flights c) Flight duration d) On-time performance }}
\label{fig:intflight}
\end{figure*}

As a case study, the Dublin-London Heathrow route has been selected to analyse in this paper. The idea is to discuss the pros and cons of replacing  aircraft flying in this route with liquid hydrogen powered ones.
The statistical data for the top 20 busiest international air routes in the world is shown in Fig.\ref{fig:intflight} and table	\ref{tab:flightint}. Fig.\ref{fig:intflight} shows the number of carried passengers (a), number of flights per year (b), flight duration (c), and \textcolor{black}{on-time performance (OTP)} (d) which refers to the level of success of the service remaining on the published schedule for each route. Table	\ref{tab:flightint} also lists the routes in the first column (IATA codes of airports are used \cite{IATA2}) followed by number of carried passengers, air distance of each route, number of flights per year, the amount of produced passenger $CO_2$/pax/leg, the duration of flight, average number of seats per aircraft, fuel burn per journey, and the OTP as an index for on time running. The data presented in this table is based on routes data in 2018 \cite{OAG}.    

\begin{table*}[!t]
	\renewcommand{\arraystretch}{1}
	\caption{Top 20 busiest international air-routes in the world}
	\label{tab:flightint}
	\centering
	\scalebox{1}{
	\begin{tabular}{c|ccccccccc}
		\hline
\# &Route 	&	Passengers 	&	Distance 	&	Flights 	&	$CO_2$	 &	Time 	&	Avg S/AC	&	Fuel &	OTP	\\ 
& 	& Million	 	&	(Km) 	&	 	&	/pax/leg (kg) &	 (minutes) 	&		&	/Journey (kg)	& 	(\%)	\\ \hline
1	&	KUL–SIN	&	4.00	&	296	&	30537	&	41.2	&	64	&	177	&	2766.3	&	72.6	\\
2	&	HKG–TPE	&	6.54	&	805	&	28887	&	81	&	107	&	282	&	8799.7	&	68	\\
3	&	CGK–SIN	&	4.66	&	880	&	27304	&	80.1	&	108	&	207	&	6328.1	&	75	\\
4	&	HKG–PVG	&	3.84	&	1255	&	21888	&	113.5	&	150	&	226	&	9551.1	&	55	\\
5	&	CGK–KUL	&	2.71	&	1127	&	19849	&	98.7	&	124	&	180	&	6197.1	&	60	\\
6	&	ICN–KIX	&	2.9	&	859	&	17488	&	79.2	&	106	&	219	&	5905.6	&	78	\\
7	&	HKG–ICN	&	3.38	&	2070	&	17075	&	154.1	&	214	&	254	&	14653.4	&	65	\\
8	&	LGA–YYZ	&	1.62	&	571	&	16956	&	92.8	&	99	&	110	&	3108.6	&	51	\\
9	&	DXB–KWI	&	2.77	&	851	&	15332	&	92	&	112	&	237	&	7283.5	&	83	\\
10	&	HKG–SIN	&	3.22	&	2562	&	15029	&	172	&	238	&	272	&	23460.5	&	70	\\
11	&	BKK–SIN	&	2.89	&	1416	&	14859	&	104.8	&	148	&	247	&	11461	&	80	\\
12	&	BKK–HKG	&	3.75	&	1687	&	14832	&	127.7	&	164	&	318	&	15074.8	&	68	\\
13	&	HKG–PEK	&	2.87	&	1989	&	14543	&	156.9	&	200	&	247	&	16225	&	57	\\
14	&	DUB–LHR	&	1.89	&	449	&	14390	&	62.5	&	90	&	165	&	2995	&	83	\\
15	&	KIX–TPE	&	2.41	&	1703	&	14186	&	138	&	197	&	225	&	12123.8	&	83	\\
16	&	JFK–LHR	&	3.05	&	5536	&	13888	&	335.5	&	418	&	264	&	54216.9	&	75	\\
17	&	KIX–PVG	&	1.77	&	1305	&	13576	&	114.5	&	168	&	174	&	8352.1	&	70	\\
18	&	ICN–NRT	&	2.27	&	1255	&	13517	&	100.7	&	140	&	211	&	8834	&	70	\\
19	&	AMS–LHR	&	1.86	&	365	&	13170	&	59.5	&	80	&	158	&	2620.2	&	75	\\
20	&	ORD–YYZ	&	1.07	&	700	&	13100	&	115.6	&	95	&	95	&	2977.2	&	63	\\		\hline
	\end{tabular}}
\end{table*}

\begin{figure}[t!]
  \centering
\includegraphics[clip = true,width=1\columnwidth,trim=0.2cm 1cm 0cm 0.1cm,]{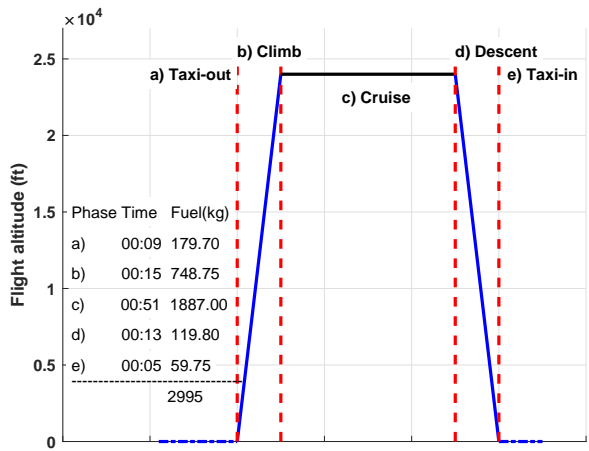}
\vspace{-3mm}
  \caption{Fuel consumption in each flight phase for DUB-LHR route }
\label{fig:flight}
\end{figure}

Fig.\ref{fig:intflight} and table	\ref{tab:flightint}, the trip between Dublin (DUB) and London Heathrow (LHR) is the busiest European entry with 14,390 flights per year according to OAG Aviation Worldwide Ltd report \cite{OAG}. DUB-LHR route is also ranked 14th in the busiest routes of the world table. The average number of flights per day would then be 40. The flights are equipped with A318, A319, A320, and A321 aircraft. Based on the last update of the OAG report the average seats/aircraft for this route is 165 that could be accommodated in the above-mentioned aircraft (total number of seats is 2328652 and 1887170 passengers (Fig.\ref{fig:intflight}.a) were carried). In Fig.\ref{fig:intflight}, the size of the circles are proportional with the average flight time in each route. The Dublin-London Heathrow average flight time (hh:mm) is 01:33 (Fig.\ref{fig:intflight}.c) and the distance is 449 km.The details of different flight segments in the selected mission is presented in Fig.\ref{fig:flight}. As it can be seen in this figure, the total fuel burn in the whole mission including taxi out, climb, cruise, descent, and taxi in phases is 2995 kg (Fig. \ref{fig:flight}). Therefore, according to International Civil Aviation Organization (ICAO) \cite{ICAO}, the aircraft fuel burn per day could be calculated as follow: 
\begin{alignat}{2}
\label{Fuel Burn}
D_f=N_f\times\gamma =&40\times2995=119.800 ton 
\end{alignat}
$N_f$ is the number of flights per day, $\gamma$ is Aircraft Fuel Burn/journey, and $D_f$ is the Daily Aircraft Fuel Burn. The total value of fuel burn per day in Dublin-London Heathrow route is 119.8 tonnes. Moreover, A318, A319, A320, A321 have a Passenger $CO_2$/pax/leg (KG) of 62.5 \cite{ICAO}. It means that the total amount of generated $CO_2$ per day is calculated as follows: 
\begin{alignat}{2}
\label{CO2}
TCO_2=&{\text{Passenger $CO_2$/pax/leg}}\times N_s \times N_f\\ =&62.5\times165 \times40=412.5 tons \notag
\end{alignat}
$N_s$ is the average seats per aircraft.

Therefore, the Dublin-London Heathrow route is generating 412.5 tonnes of $CO_2$ per day. This emission could be cut with liquid hydrogen-powered aircraft subject to a feasible and affordable procedure of producing hydrogen fuel in quantity from a low-carbon sources. 
The heating value of liquid hydrogen is 2.8 times higher than jet fuel. Therefore, the daily need of the liquid hydrogen for the route is 119800/2.8 = 42.785 tonnes. table \ref{tab:flight} compares the required fuel for jet fuel powered aircraft with those of hydrogen fuel powered aircraft for the selected route per day. 

\begin{table}[!t]
	\renewcommand{\arraystretch}{1.3}
	\caption{Fuel required for Dublin-London Heathrow flights per day}
	\label{tab:flight}
	\centering
	\scalebox{0.9}{
	\begin{tabular}{c|cccc}
		\hline
Aircraft Type	& $N_f$	&FHV (MJ/kg) &	$\gamma$ (kg)	&$D_f$ (kg)\\ \hline
Jet Fuel Powered &40 & 43.1 &2995 &119800 \\
Hydrogen Fuel Powered &	40&	120&	1070&	42785 \\
		\hline
	\end{tabular}}
\end{table}

The Clean Energy Partnership (CEP) states that producing hydrogen by electrolysis requires about 55 kWh/kg $H_2$ of electricity at an assumed rate of efficiency of higher than 60 percent \cite{CEP}. The daily required electric energy to produce the required liquid hydrogen for the busiest European route (Dublin-London-Heathrow) is: 
$D_{H_2}=42785\times55$ = 2353.1 MWh.  
The investment cost ($\Xi$) for P2H plant is assumed to be \euro236000/MW \cite{Nrel}. 

\subsection{Transmission network data}
\begin{figure}[ht]
	\centering
		\includegraphics[width=0.2\columnwidth,bb=250 300 350 700]{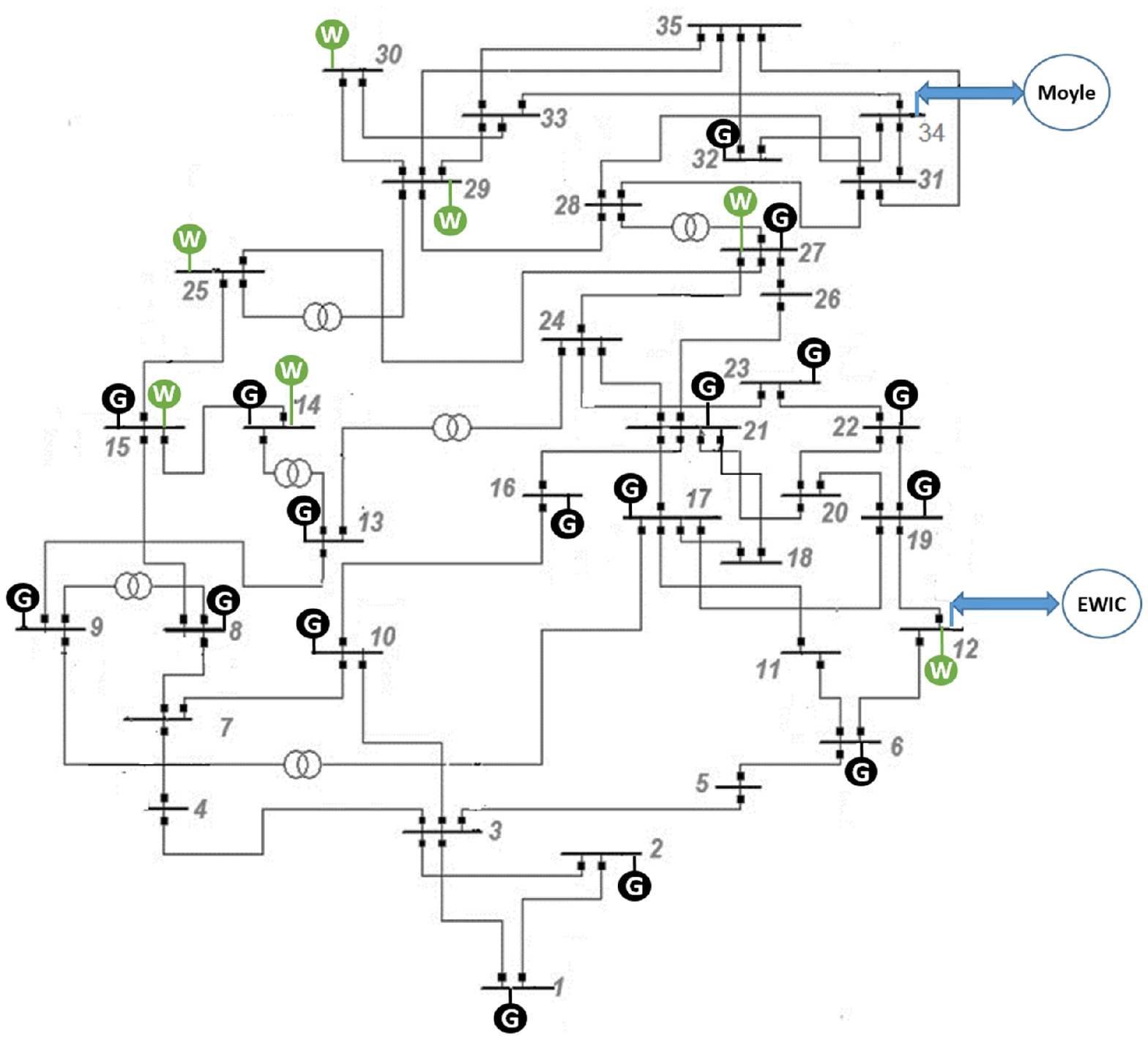}
\caption{Irish representative transmission network}
	\label{fig:irishnetwork}
\end{figure}

The transmission line data of Irish network is taken from \cite{5643190} as shown in Fig.\ref{fig:irishnetwork}.  

\begin{figure}[t!]
  \centering
		\includegraphics[width=0.19\columnwidth,bb=265 290 370 500]{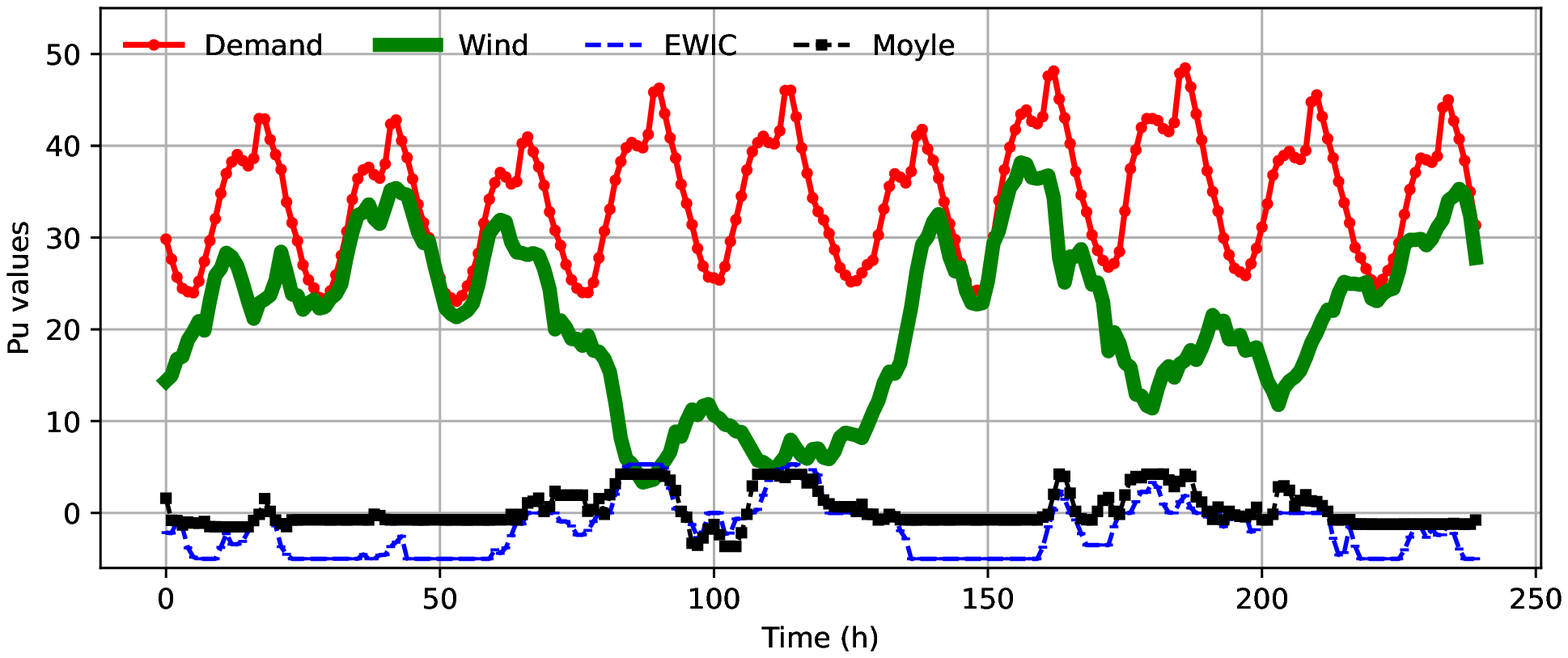}
  \caption{Demand, wind and import/export variation with time}
\label{fig:irishdata}
\end{figure}

\begin{table}[!t]
	\renewcommand{\arraystretch}{1.3}
	\caption{Generation data and connection point}
	\label{tab:gencapcon}
	\centering
	\scalebox{0.8}{
	\begin{tabular}{ccc|ccc}
		\hline
Generator	&	Cap (MW)	&	b	&	Generator	&	Cap (MW)	&	b	\\ \hline
$g_1$	&	90	&	1	&	$g_{17}$	&	104	&	27	\\
$g_2$	&	90	&	1	&	$g_{18}$	&	230	&	22	\\
$g_3$	&	431	&	1	&	$g_{19}$	&	230	&	22	\\
$g_4$	&	405	&	22	&	$g_{20}$	&	52	&	16	\\
$g_5$	&	61	&	19	&	$g_{21}$	&	52	&	16	\\
$g_6$	&	118	&	17	&	$g_{22}$	&	81	&	15	\\
$g_7$	&	58	&	17	&	$g_{23}$	&	81	&	15	\\
$g_8$	&	58	&	17	&	$g_{24}$	&	54	&	9	\\
$g_9$	&	431	&	6	&	$g_{25}$	&	54	&	9	\\
$g_{10}$	&	342	&	23	&	$g_{26}$	&	241	&	9	\\
$g_{11}$	&	408	&	23	&	$g_{27}$	&	241	&	9	\\
$g_{12}$	&	17	&	32	&	$g_{28}$	&	52	&	10	\\
$g_{13}$	&	91	&	21	&	$g_{29}$	&	52	&	10	\\
$g_{14}$	&	285	&	14	&	$g_{30}$	&	400	&	8	\\
$g_{15}$	&	285	&	14	&	$g_{31}$	&	137	&	13	\\
$g_{16}$	&	285	&	14	&	$g_{32}$	&	444	&	2	\\
		\hline
	\end{tabular}}
	\label{tab:gendata}
\end{table}
The capacity of each conventional generating unit as well as the connection bus is specified in table \ref{tab:gendata}.
The peak values for the demand nodes are provided in table \ref{tab:loaddata}. The peak demand for the whole network is 5400 MW.
\begin{table}[!t]
	\renewcommand{\arraystretch}{1.3}
	\caption{Forecasted values for peak demand  in each bus (MW).}
	\label{tab:loaddata}
	\centering
	\scalebox{0.9}{
	\begin{tabular}{cc|cc|cc}
		\hline
b	&	$\bar{P}^L_b$	&	b	&	$\bar{P}^L_b$	&	b	&	$\bar{P}^L_b$	\\
\hline
1	&	175.65	&	12	&	144.65	&	25	&	242.81	\\
2	&	7.75	&	13	&	15.07	&	26	&	224.74	\\
3	&	224.74	&	15	&	269.50	&	27	&	219.56	\\
4	&	61.12	&	16	&	188.55	&	28	&	292.75	\\
5	&	220.42	&	17	&	51.66	&	29	&	103.31	\\
6	&	28.31	&	19	&	348.70	&	30	&	256.57	\\
7	&	110.21	&	20	&	346.98	&	31	&	222.99	\\
9	&	74.05	&	21	&	229.03	&	32	&	172.19	\\
10	&	190.30	&	22	&	567.41	&	33	&	124.85	\\
11	&	87.22	&	23	&	60.27	&	35	&	138.64	\\
		\hline
	\end{tabular}}
	\label{Forecasting output}
\end{table}
The SNSP is assumed to be 70\%. 
It is assumed that there are 7 wind turbine sites in the network (with total installed capacity of 4200 MW. The installed capacity of each wind farm is given in table \ref{tab:winddata}.
\begin{table}[!t]
	\renewcommand{\arraystretch}{1.3}
	\caption{Installed capacity of wind turbines  in each bus (MW).}
	\label{tab:winddata}
	\centering
	\scalebox{0.9}{
	\begin{tabular}{ccccccc}
		\hline
$w_1$	& $w_2$	&$w_3$&	$w_4$	&$w_5$	&$w_6$&	$w_7$\\ \hline 
b=12 &b=14 & b=15 &b=25 &b=27 &b=29 &b=30 \\
611 &	648&	666&	537&	629&	537&	574\\
		\hline
	\end{tabular}}
\end{table}
for this study, the realistic demand, wind and import/export variation with time is taken from EirGrid \cite{EirGrid} as shown in Fig. \ref{fig:irishdata}. The data is for 10 days (240 hours). 

\subsection{No P2H case}
In this case, we assume that there is no P2H plant in the system. The daily quantities calculated in this case are:   
\begin{itemize}
    \item Daily Costs=\euro6.910M 
    \item Daily WC=4.1586 GWh. 
    \item Daily $CO_2$ in power system=14593.55 tons 
    \item Daily $CO_2$ in aviation= \textcolor{black}{412.5 tones}
\end{itemize}
\subsection{Base case) P2H at Dublin airport}
In this case, it is assumed that the P2H plant is installed at bus 22 (Dublin). The problem is solved and the simulation results show that
\begin{itemize}
    \item Daily Costs= \euro6.610M
    \item Daily WC= 2.020 GWh
    \item Daily $CO_2$ in power system= 14697.84 tons
    \item Daily $CO_2$ in aviation= 0 tones 
\end{itemize}
This means that we can have some saving by reducing the wind curtailment costs as well as 308.21 tons drop in total daily $CO_2$ pollution. 
The optimal size of the P2H power plant is $\xi=504.255$ MW.

\begin{figure}[th]
  \centering
	\includegraphics[width=0.5\columnwidth,bb=120 10 340 290]{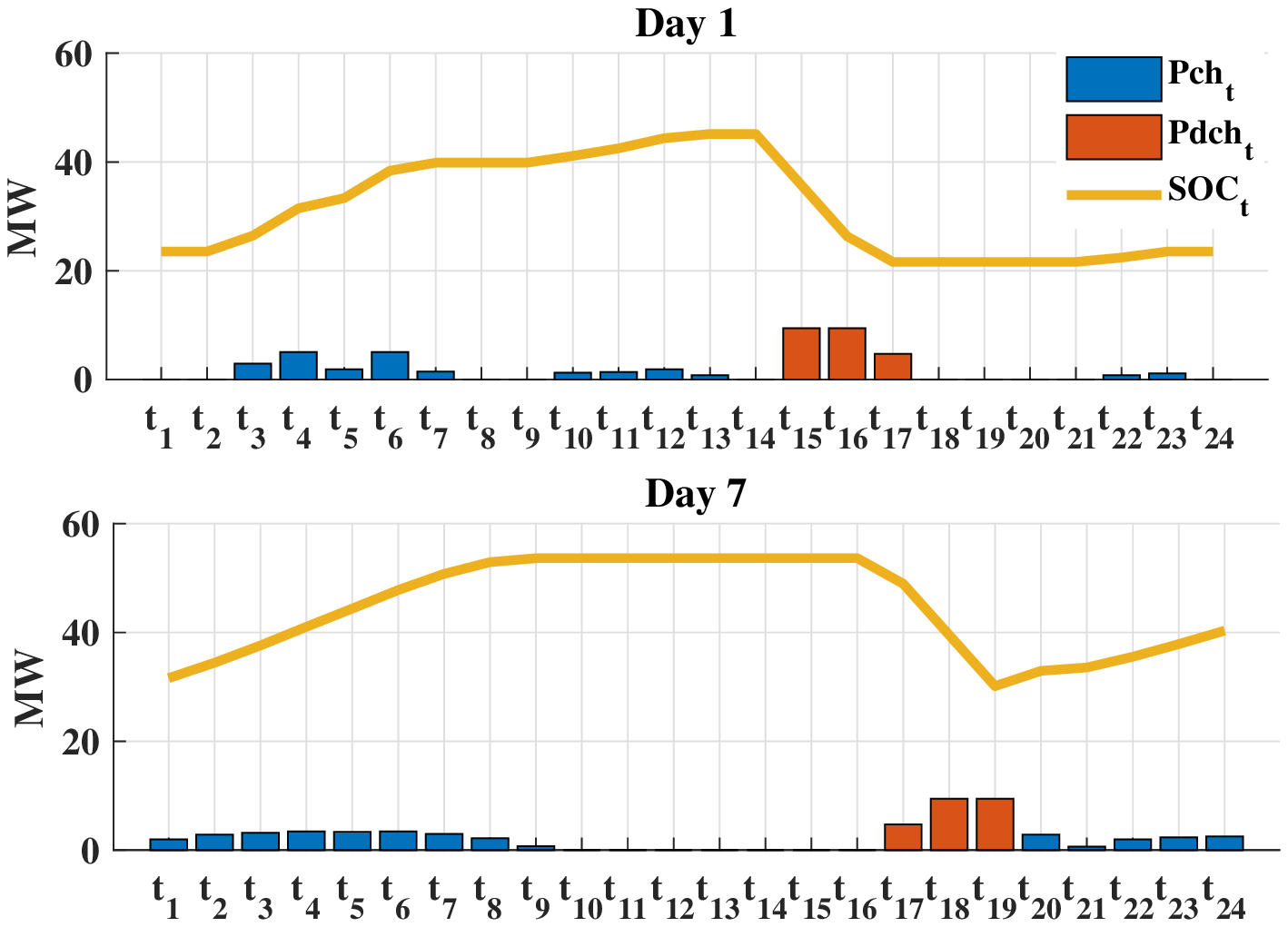}
  \caption{P2H hourly schedule for base case for some selected days}
\label{fig:schedule}
\end{figure}
The hourly schedule of P2H for the base case is shown in Fig. \ref{fig:schedule} for some selected days. \textcolor{black}{The optimal charge and discharge pattern of the hydrogen storage is depicted in this figure. The hourly variation of state of charge ($SOC_t$) shows the available amount of hydrogen at each time step. Whenever the hydrogen storage is charged then the magnitude of $SOC_t$ increases. The shapes of charging/discharging patterns are affected by electric demand, wind generation and network characteristics.For example, if a sufficient amount of wind is available in the network then the charging of hydrogen storage will begin}. 

\subsection{Sensitivity analysis}
\subsubsection{Connection point}
\begin{figure}[t!]
  \centering
	\includegraphics[width=0.5\columnwidth,bb=160 15 480 290]{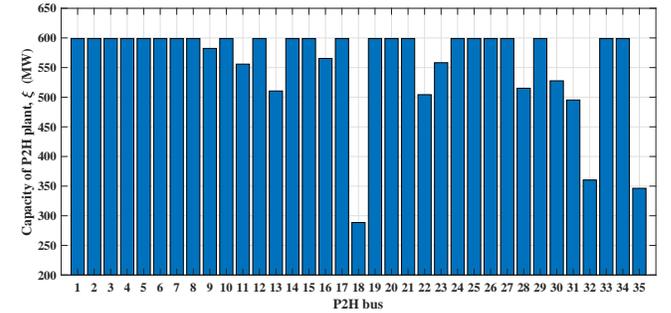}
    \caption{Impact of P2H plant location on its optimal capacity}
\label{fig:P2hcap}
\end{figure}

Assuming that there is only one P2H plant in the system, the connection point of P2H plant to the electricity grid is varied from bus 1 to bus 35. 
The optimal capacity of the P2H plant is determined in each case. The results are shown in Fig. \ref{fig:P2hcap}. The minimum value of $\xi$ is 288 MW at bus 18 and the maximum 600 MW on bus 15. 

\begin{figure}[t!]
  \centering
	\includegraphics[width=0.8\columnwidth,bb=110 295 500 500]{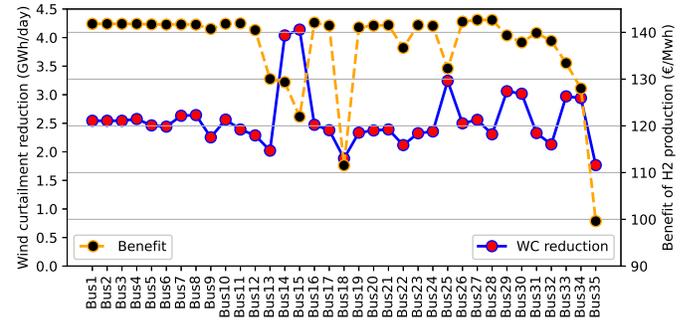}
    \caption{\textcolor{black}{Impact of P2H plant location on economic benefits and wind curtailments}}
\label{fig:wcred}
\end{figure}
The price of producing $H_2$ is highly dependent on the location of the P2H plant as shown in Fig. \ref{fig:wcred}. 
The benefit of producing $H_2$ varies from 99.36 to 142.46 \euro/MWh. It means that producing $H_2$ is helping the system to avoid other costs (i.e. wind curtailment subsidies and environmental penalties). However, it should be noted that these numbers will not necessarily remain constant if the total daily required $H_2$ changes. 
\textcolor{black}{
\textcolor{black}{It is worth considering that even the worst-case scenario in Fig. \ref{fig:wcred} (bus 35) would be much more beneficial compared to conventional jet fuel. The price of jet fuel, $P_{f}$, is around \euro0.5/kg \cite{IATA}. Moreover, the carbon offsetting cost, $P_{cos}$, in different schemes would vary between \euro0.2/kg and \euro0.36/kg \cite{Offsett1,Offsett2}. Therefore, based on table \ref{tab:flight},  the minimum and maximum equivalent jet fuel cost, $P_{ef}$, would be  25.45\euro/MWh (no carbon offsetting cost) and 43.86\euro/MWh (maximum carbon offsetting cost) respectively as described in (\ref{eq:efuelprice}). All scenarios in Fig. \ref{fig:wcred} will produce hydrogen fuel cheaper than jet fuel for the airliners.}}
\textcolor{black}{
\begin{alignat}{2}
\label{eq:efuelprice}
P_{ef}= \frac{D_{f}(P_{f}+P_{cos})}{D_{H_2}}
\end{alignat}
}

\textcolor{black}{$P_{ef}$ is the equivalent jet fuel price per MWh, $P_{f}$ is the fuel price, and $P_{cos}$ is the carbon offsetting cost.} 

One of the factors that makes the P2H economically viable is the ability to absorb more wind to the electric network (and producing $H_2$). However, this capability in reducing the wind curtailment is highly dependant on the connection point of the P2H plant. Fig. \ref{fig:wcred} shows the Wind curtailment reduction (GWh) vs the connection point of the P2H plant. 
It varies between 1.767 GWh (bus 35) to 4.139 GWh (bus 15). 

\subsubsection{SNSP level}

\begin{figure}[t!]
  \centering
		\includegraphics[width=0.17\columnwidth,bb=245 160 340 580]{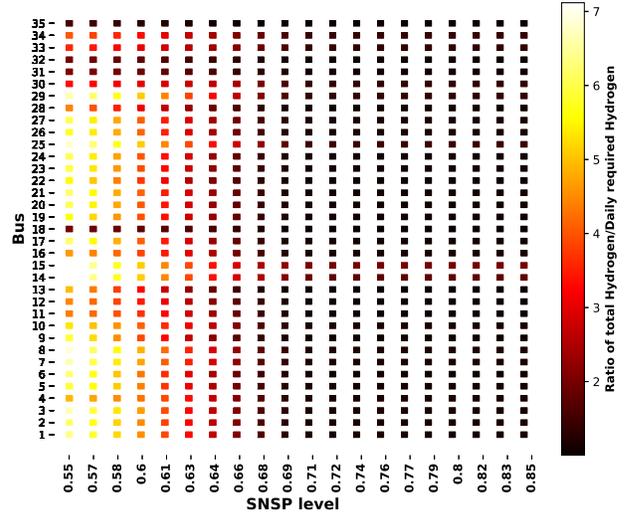}
  \caption{\textcolor{black}{Impact of SNSP level on capability of generating $H_2$}}
\label{fig:SNSP}
\end{figure}

Finally, the impact of the SNSP level \eqref{eq:SNSP} is investigated on the total absorbed $H_2$. For this purpose, the SNSP level is changed from 0.55 to 0.8. As depicted in Fig.\ref{fig:SNSP}, the SNSP level has a direct impact on total producible $H_2$. This impact is different if the connection point of the P2H plant is changed in the network. For example, Bus 15 can absorb up to 7 times the daily required $H_2$ (2353.1 MWh) at a low SNSP level (0.55). It is expected to happen since at low SNSP levels, the total wind curtailment increases and it is more economic to convert it to hydrogen. \textcolor{black}{This clearly demonstrates the importance of optimal allocation of P2H plant in the electricity network.}

From Fig. \ref{fig:SNSP} it is observed that increasing the SNSP beyond a certain level (here 75\%) does not lead to increase in $H_2$ production or even reducing the operating costs. The technical reason behind this phenomena is that after certain SNSP level, the system can not absorb more power due to transmission line thermal limits. The min power generation limit of other non-res technologies can be also a reason for it. 
\subsubsection{Selecting a pair of buses for P2H connection}
\begin{figure}[t!]
  \centering
	\includegraphics[width=0.27\columnwidth,bb=180 0 300 400]{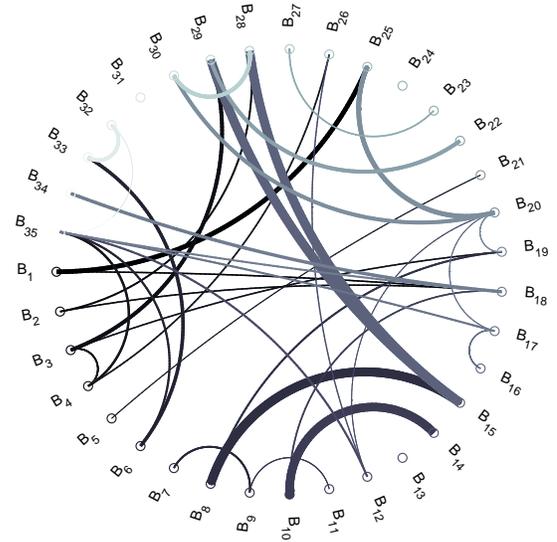}
  \caption{Impact of pair-buses for connecting the P2H plants in reducing the wind curtailment}
\label{fig:pair}
\end{figure}

Increasing the number of P2H plants can improve the capability of the system in absorbing wind and therefore reducing the wind curtailment. The analysis shows that the best buses for pairing are $b_{15}-b_{8}$ which can reduce the initial average daily wind curtailment by 4.155 GWh. Both of these buses host wind turbines and therefore this combination has the potential to absorb more wind from the grid. The weight of edges in Fig. \ref{fig:pair}, shows the merit of each pair of buses for reducing the wind curtailment. 

\begin{figure}[t!]
  \centering
	\includegraphics[width=0.2\columnwidth,bb=180 15 263 370]{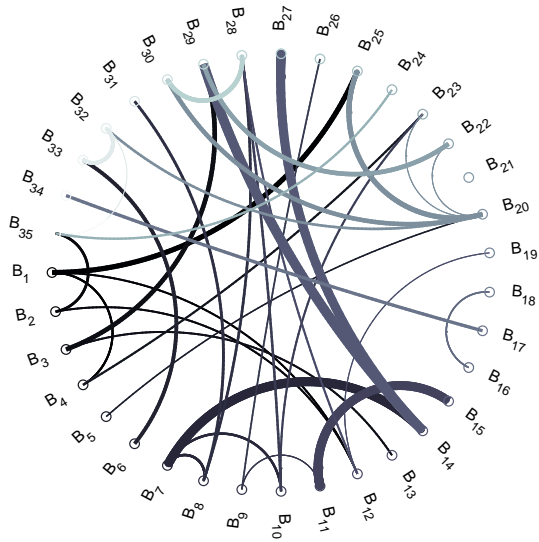}
  \caption{Impact of pair-buses for connecting the P2H plants in reducing the operating costs}
\label{fig:paircost}
\end{figure}

It should be noted that if the purpose of having two P2H plants is reducing the operating costs then the ''best pair'' might be different. Fig. \ref{fig:paircost} shows the best combination of the buses for reducing the operating costs. The best pair is $b_{14}-b_{7}$ which can reduce the daily operating costs (compared to No $H_2$ case) by \euro 0.6261M. 

\subsection{Airlines' perspective}
The possibility of having hydrogen fuel-powered civil aircraft (100+ passenger planes) is a game-changer for airliners; especially for dealing with the future targets and requirements of air travel systems (e.g. Flight Path 2050 \cite{ACARE}). The majority of a flight mission in civil aircraft is associated with the cruise phase at high altitudes where emissions have two to four times the impact of equivalent emissions at ground level. \\
\indent Moreover, producing an alternative sustainable fuel in quantity with affordable prices in comparison with jet fuels is a strong motivation for aero-engine designers and manufacturers to invest and explore new designs for the next generation of aircraft engines respect to the limited supply and increasing price of the current jet fuels. In this aspect, electrification is also being considered as a high potential candidate. There are some successful projects in small battery-powered aircraft development \cite{Electric1,Electric2,Electric3}. However, the limitation of batteries' weight and low power to weight ratio means that it will be difficult to scale up to larger aircraft. 

“With hydrogen as a fuel, there is no physical reason we can’t go larger and longer. Each step in increasing the range and the size of aircraft we can fly needs technological improvements, but it’s mostly a question of engineering, not new physics,” Miftakhov, who previously founded an electric vehicle charging company eMotorWerks, said \cite{speechh}.
Although engine and aircraft manufacturers have to redesign the aircraft and engine, it is not a crucial point as it is what has happened every time a new engine is introduced. 

\section{Conclusion}
\label{sec:Conclusion}
The main findings of this paper are outlined as follows:
\begin{itemize}
    \item \textcolor{black}{The connection point of the P2H plant as well as the size of it have significant impacts on the capability of P2H in absorbing electricity. This is mainly because of transmission network physical constraints (power flow and thermal line limits).} 
    \item \textcolor{black}{By having P2H plants close to the wind sites, it would be easier to absorb the generated clean wind power and avoid violating the line flow limits}.
    \item The environmental pollution penalty can make the P2H economically viable. \textcolor{black}{The generated hydrogen can be either blended with the natural gas or combusted directly with nearly zero carbon emission}.
    \item Producing the $H_2$ in high penetrated wind systems can be done at a negative price. \textcolor{black}{The income obtained from selling the hydrogen and offsetting the carbon emission can return the operation and investment costs}.
    \item \textcolor{black}{Converting the excessive wind using the P2G can reduce the dependency rate on energy imports. It is inline with the stability improvement of EU’s energy supply.
    }
    \item The idea of producing green hydrogen could help the airliners in ACARE Flight Path 2050 emission requirements satisfaction as the aircraft greenhouse gases will be reduced dramatically.\textcolor{black}{Using hydrogen ensures affordable transportation for all consumers by reducing the operating costs of airlines in the long run.}
\end{itemize}
\textcolor{black}{
Suggestions for future work:
\begin{itemize}
    \item The impact of $H_2$ extraction on gas networks should be investigated. A portion of the electricity demand will be supplied via thermal power plants. These generators are supplied by gas networks. The technical and economic set-points of gas networks will be affected \cite{8826763}. \item A more detailed AC-OPF can better characterize the impact of $H_2$ extraction in power systems. 
    \item The voltage stability of power system in presence of $H_2$ extraction plants should be investigated \cite{rabiee2018information}. The available transfer capacity of the system will be different in this case. 
    \item The uncertainty of wind power generation should be taken into account to avoid financial and technical risks \cite{HEMMATI202041}. 
\end{itemize}
}
\bibliographystyle{IEEEtran}
\bibliography{ref}

\end{document}